\documentclass{article}
\usepackage{graphicx,lscape}

\begin{document}
\frenchspacing

\title{MOST: detecting cancer differential gene expression}

\author{HENG LIAN} 

\maketitle
\pagestyle{headings}


\setcounter{footnote}{1}

\begin{abstract}
{We propose a new statistics for the detection of differentially expressed genes, when the genes are activated only in a subset of the samples. Statistics designed for this unconventional circumstance has proved to be valuable for most cancer studies, where oncogenes are activated for a small number of disease samples. Previous efforts made in this direction include COPA (\cite{tomlins2005}), OS (\cite{tib2006}) and ORT (\cite{wu2007}). We propose a new statistics called maximum ordered subset t-statistics (MOST) which seems to be natural when the number of activated samples is unknown. We compare MOST to other statistics and find the proposed method often has more power then its competitors.}
{Cancer; COPA; Differential gene expression; 
Microarray.}\vspace*{-6pt}
\end{abstract}

\section{Introduction}
The most popular method for differential gene expression detection in two-sample microarray studies is to compute the t-statistics. The differentially expressed genes are those whose t-statistics exceed a certain threshold. Recently, due to the realization that in many cancer studies, many genes show increased expressions in disease samples, but only for a small number of those samples. The study of \cite{tomlins2005} shows that t-statistics has low power in this case, and they introduced the so-called ``cancer outlier profile analysis" (COPA). Their study shows clearly that COPA can perform better than the traditional t-statistics for cancer microarray data sets.
 
More recently, several progresses have been made in this direction with the aim to design better statistics to account for the heterogeneous activation pattern of the cancer genes. In \cite{tib2006}, the authors introduced a new statistics, which  they called outlier sum. Later, \cite{wu2007} proposed outlier robust t-statistics (ORT) and showed it usually outperformed the previously proposed ones in both simulation study and application to real data set.

In this paper, we propose another statistics for the detection of cancer  differential gene expression which have similar power to ORT when the number of activated samples are very small, but perform betters when more samples are differentially expressed. We call our new method the maximum ordered subset t-statistics (MOST). Through simulation studies we found the new statistics outperformed the previously proposed ones under some circumstances and never significantly worse in all situations. Thus we think it is a valuable addition to the dictionary of cancer outlier expression detection.

\section{\label{sec.def}Maximum ordered subset t-statistics (MOST)}
We consider the simple 2-class microarray data for detecting cancer genes. We assume there are $n$ normal samples and $m$ cancer samples. The gene expressions for normal samples are denoted by $x_{ij}$ for genes $i=1,2,\ldots,p$ and samples $j=1,2,\ldots n$, while $y_{ij}$ denote the expressions for cancer samples with $i=1,2,\ldots,p$ and $j=1,2,\ldots m$. In this paper, we are only interested in one-sided test where the activated genes from cancer samples have a higher expression level. The extension to two-sided test is straightforward.

The usual t-statistics (up to a multiplication factor independent of genes) for two-sample test of differences in means is defined for each gene $i$ by
\begin{equation}\label{t}
T_i=\frac{\bar{x}_i-\bar{y}_i}{s_i},
\end{equation}
where $\bar{x}_i=\sum_jx_{ij}/n$ is the average expression of gene $i$ in normal samples,  $\bar{y}_i=\sum_jy_{ij}/m$ is the average expression of gene $i$ in cancer samples, and $s_i$ is the usual pooled standard deviation estimate
\[
s_i^2=\frac{\sum_{1\le j\le n}(x_{ij}-\bar{x}_i)^2+\sum_{1\le j\le m}(y_{ij}-\bar{y}_i)^2}{n+m-2}.
\]
The t-statistics is powerful when the alternative distribution is such that $y_{ij}, j=1,2,\ldots,m$ all come from a distribution with a higher mean. \cite{tomlins2005} argues that for most cancer types, heterogeneous activation patterns  make t-statistics inefficient for detecting those expression profiles. They defined the COPA statistics 
\begin{equation}\label{copa}
C_i=\frac{q_r(\{y_{ij}\}_{1\le j\le m})-med_i}{mad_i},
\end{equation}
where $q_r(\cdot)$ is the $r$th percentile of the data, $med_i=median(\{x_{ij}\}_{1\le j\le n},\{y_{ij}\}_{1\le j\le m})$ is the median of the pooled samples for gene $i$, and $mad_i=1.4826\times median(\{x_{ij}-med_i\}_{1\le j\le n},\{y_{ij}-med_i\}_{1\le j\le m})$ is the median absolute deviation of the pooled samples.

The choice of $r$ in (\ref{copa}) depends on the subjective judgement of the user. The use of $med_i$ and $mad_i$ to replace the mean and the standard deviation in (\ref{t}) is due to robustness considerations since it is already known that some of the genes are differentially expressed.

In (\ref{copa}), only one value of $\{y_{ij}\}$ is used in the computation. A more efficient strategy would be to use additional expression values. Let 
\begin{equation}\label{outlier}
O_i=\{y_{ij}:y_{ij}>q_{75}(\{x_{ij}\}_{1\le j\le n},\{y_{ij}\}_{1\le j\le m})+IQR(\{x_{ij}\}_{1\le j\le n},\{y_{ij}\}_{1\le j\le m})\}
\end{equation}
be the outliers from the cancer samples for gene $i$, where $IQR(\cdot)$ is the interquartile range of the data. The OS statistics from \cite{tib2006} is then defined as
\begin{equation}\label{os}
OS_i=\frac{\sum_{y_{ij}\in O_i}(y_{ij}-med_i)}{mad_i}.
\end{equation}
More recently, \cite{wu2007} studied ORT statistics, which is similar to OS statistics. The important difference that makes ORT superior is that outliers are defined relative to the normal sample instead of the pooled sample. So in their definition,
\begin{equation}\label{os}
O_i=\{y_{ij}:y_{ij}>q_{75}(\{x_{ij}\}_{1\le j\le n})+IQR(\{x_{ij}\}_{1\le j\le n})\}
\end{equation}
By similar reasoning $med_i$ in OS is replaced by $med_{ix}$ and $mad_j$ by $median(\{x_{ij}-med_{ix}\}_{1\le j\le n},\{y_{ij}-med_{iy}\}_{1\le j\le m})$, where $med_{ix}$ and $med_{iy}$ are the medians of normal and cancer samples respectively.

In both OS and ORT statistics, the outliers are defined somewhat arbitrarily with no convincing reasons. To address this question, we propose the following statistics that implicitly considers all possible values for outlier thresholds. 

Suppose for notational simplicity that $\{y_{ij}\}_{1\le j\le m}$ are ordered for each $i$: 
\[
y_{i1}\ge y_{i2}\ge \cdots\ge y_{im}.
\]
If the number of samples where oncogenes are activated were known, we would naturally define the statistics as
\begin{equation}
M_{ik}=\frac{\sum_{1\le j\le k}(y_{ij}-med_{ix})}{median(\{x_{ij}-med_{ix}\}_{1\le j\le n},\{y_{ij}-med_{iy}\}_{1\le j\le m})}.
\end{equation}
When $k$ is not known to us, one would be tempted to define 
\[
M_i=\max_{1\le k\le m}M_{ik}.
\]
But this does not quite work since obviously $M_{ik}$ for different values of $k$ are not directly comparable under the null distribution that $x_{ij},y_{ij}\sim N(0,1)$. For example, when $m=2$, we have $E[y_{i1}-med_{ix}]>0$ while $E[\sum_{j=1,2} (y_{ij}-med_{ix})]=0$. This observation motivates us to normalize $M_{ik}$ such that each approximately has mean $0$ and variance $1$. This can be achieved by defining $\mu_k=E[\sum_{1\le j\le k}z_{j}]$ and $\sigma_k^2=Var(\sum_{1\le j\le k}z_{j})$ where $z_1>z_2>\cdots >z_m$ is the order statistics of $m$ samples generated from the standard normal distribution. Then we can define $M_{ik}$ as:
\begin{equation}
M_{ik}=\left(\frac{\sum_{1\le j\le k}(y_{ij}-med_{ix})}{1.4826\times median(\{x_{ij}-med_{ix}\}_{1\le j\le n},\{y_{ij}-med_{iy}\}_{1\le j\le m})}-\mu_k\right)/\sigma_k,
\end{equation}
so that $M_{ik}$ has mean and variance approximately equal to $0$ and $1$ respectively. 

Finally we can define our new statistics (called MOST) as
\begin{equation}
M_i=\max_{1\le k\le m}M_{ik}.
\end{equation}
With MOST, we practically consider every possible threshold above which $y_{ij}$ are taken to be outliers. In this formulation, the number of outliers is implicitly defined as
\begin{equation}\label{k}
\arg\max_{1\le k\le m}M_{ik}.
\end{equation}

\begin{figure}
\centering
\begin{tabular}{cc}
\includegraphics[width=6.5cm]{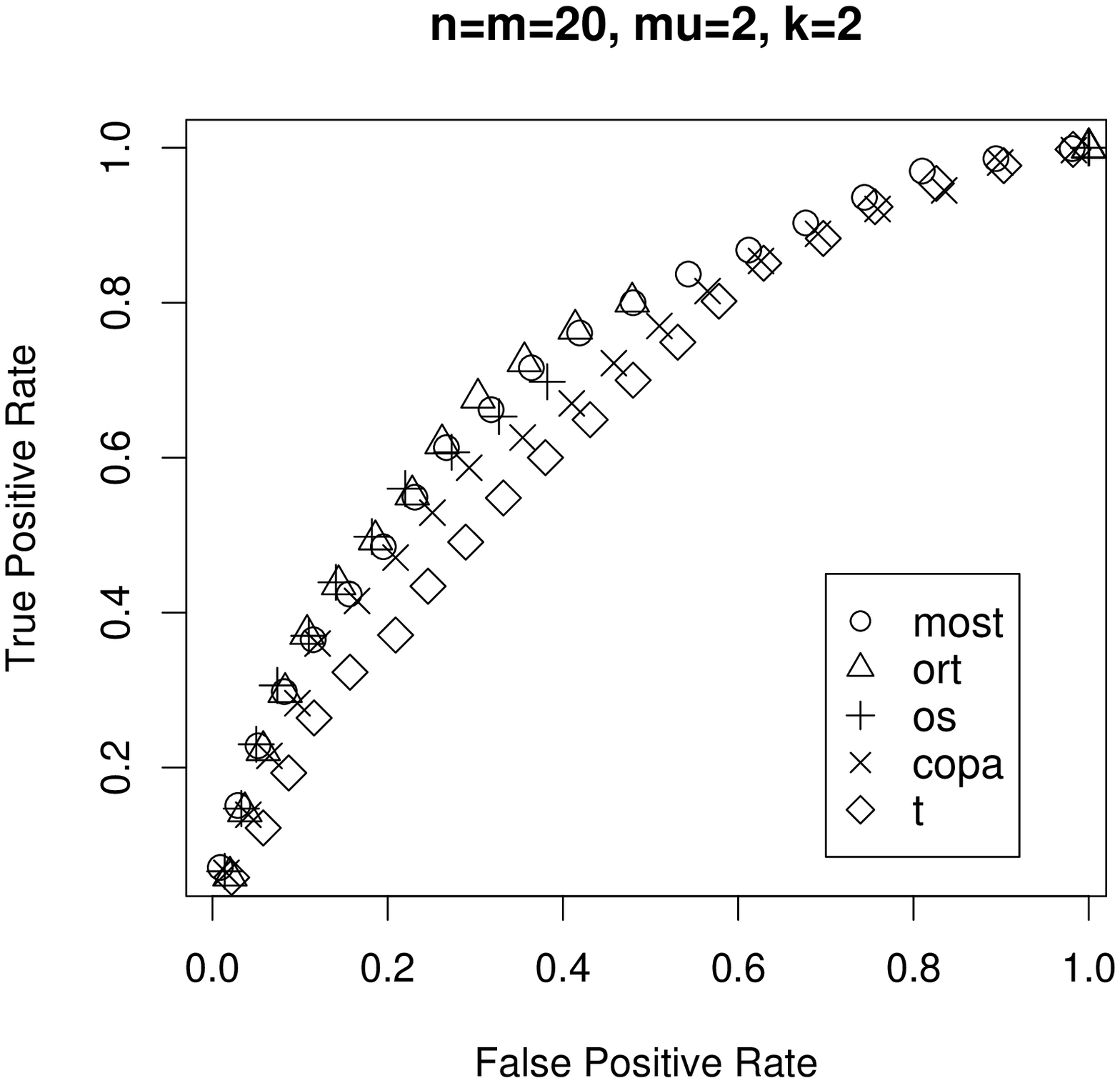} &
\includegraphics[width=6.5cm]{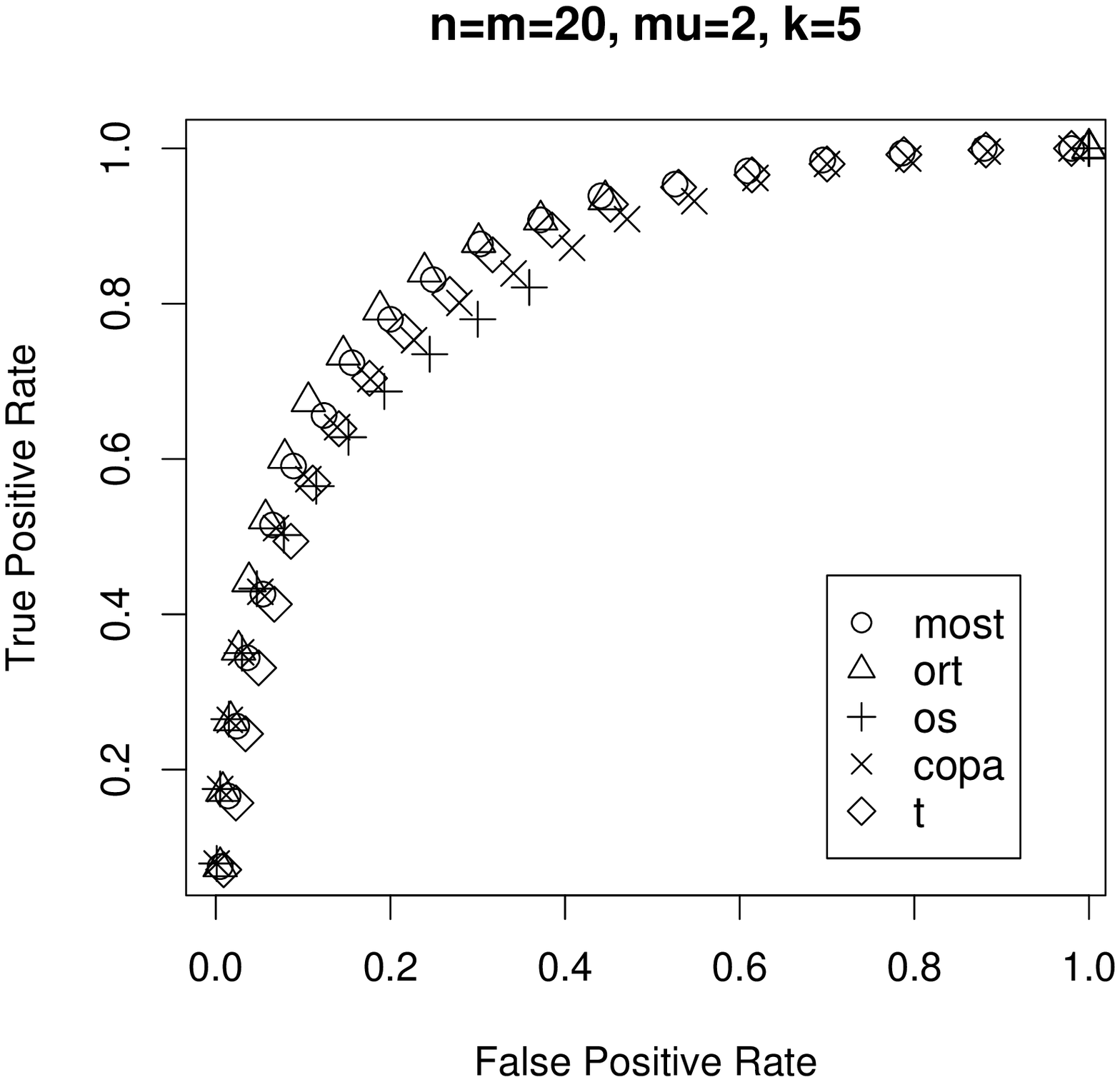} \\
\includegraphics[width=6.5cm]{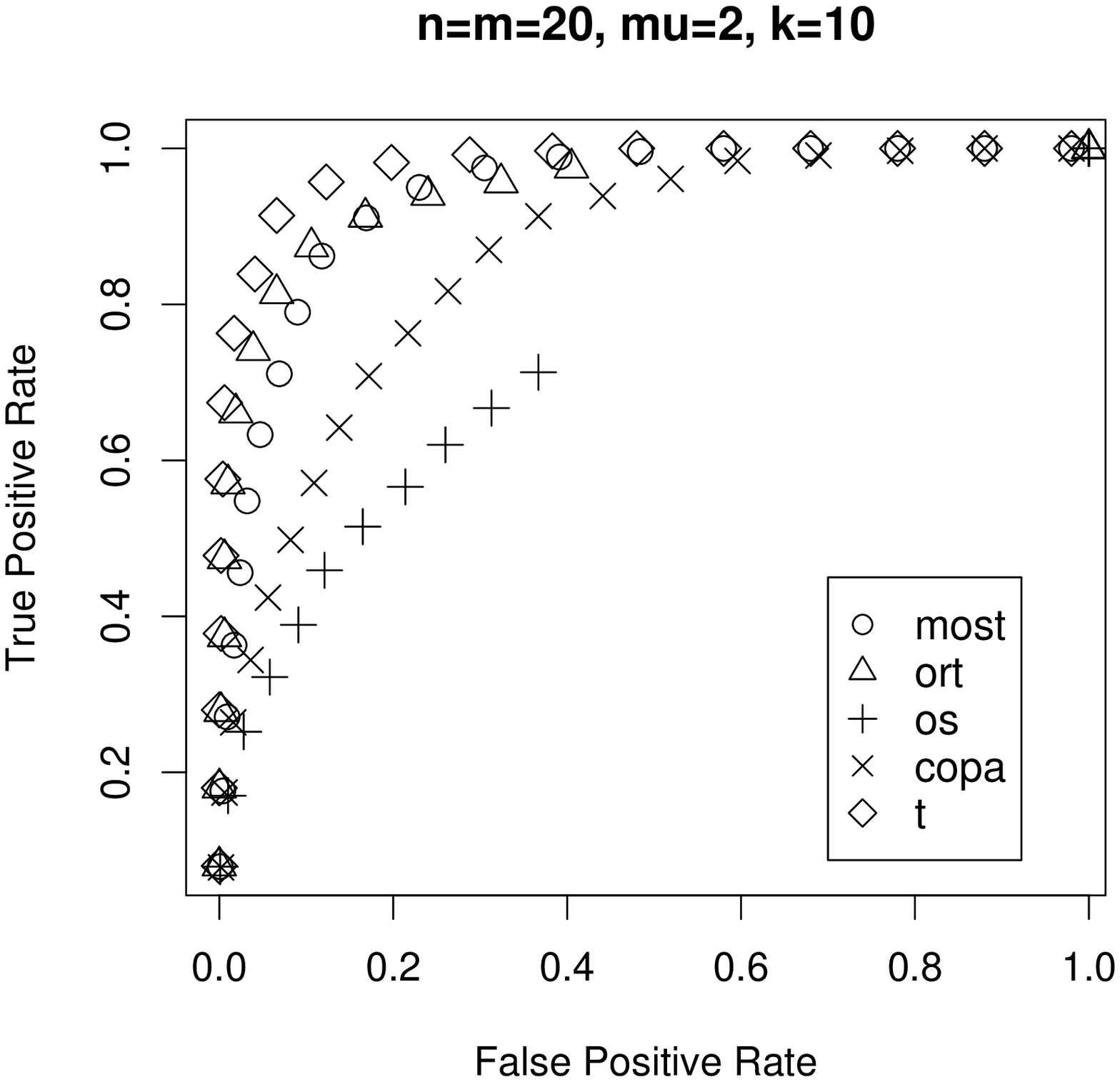} &
\includegraphics[width=6.5cm]{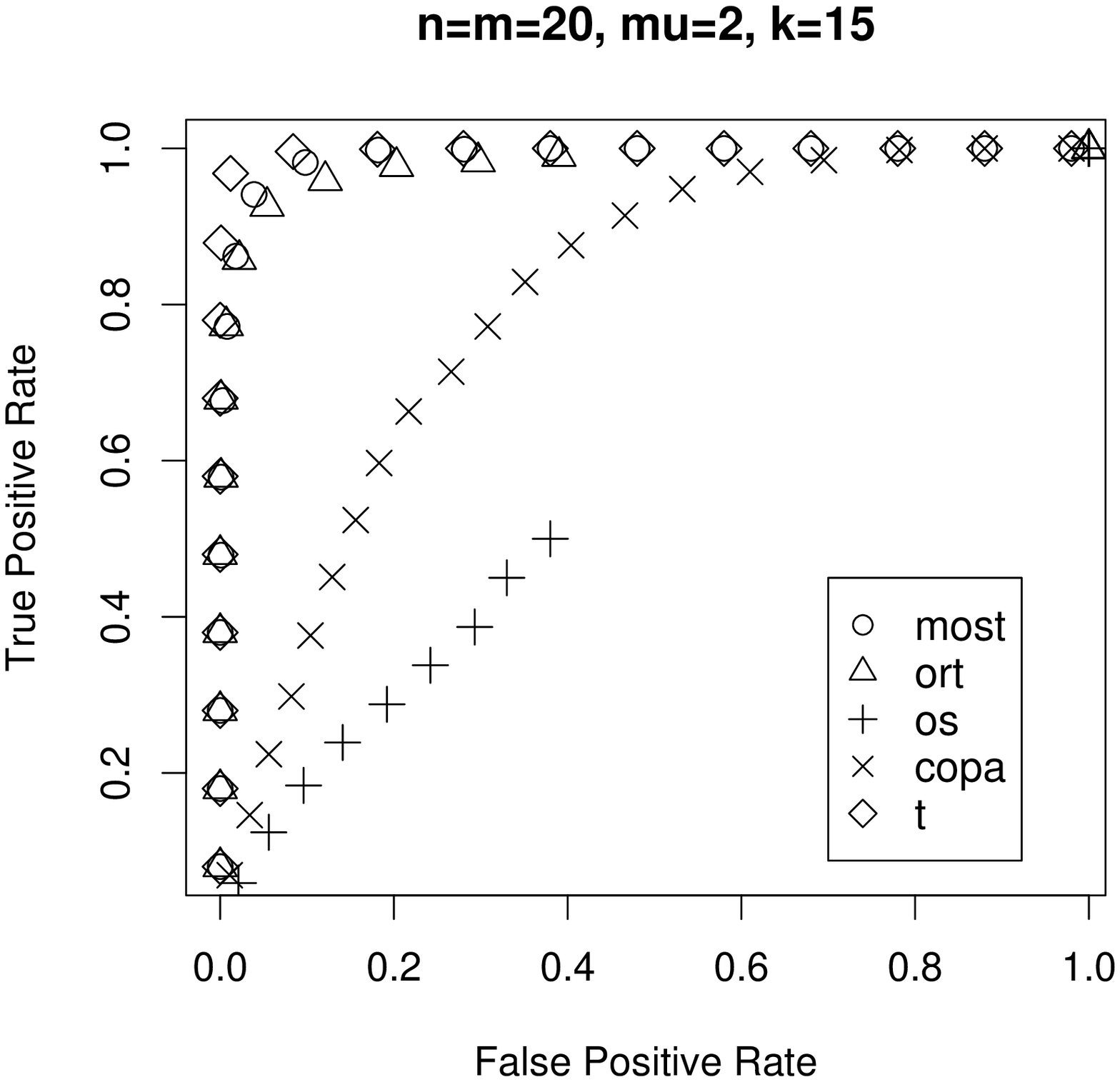} \\
\includegraphics[width=6.5cm]{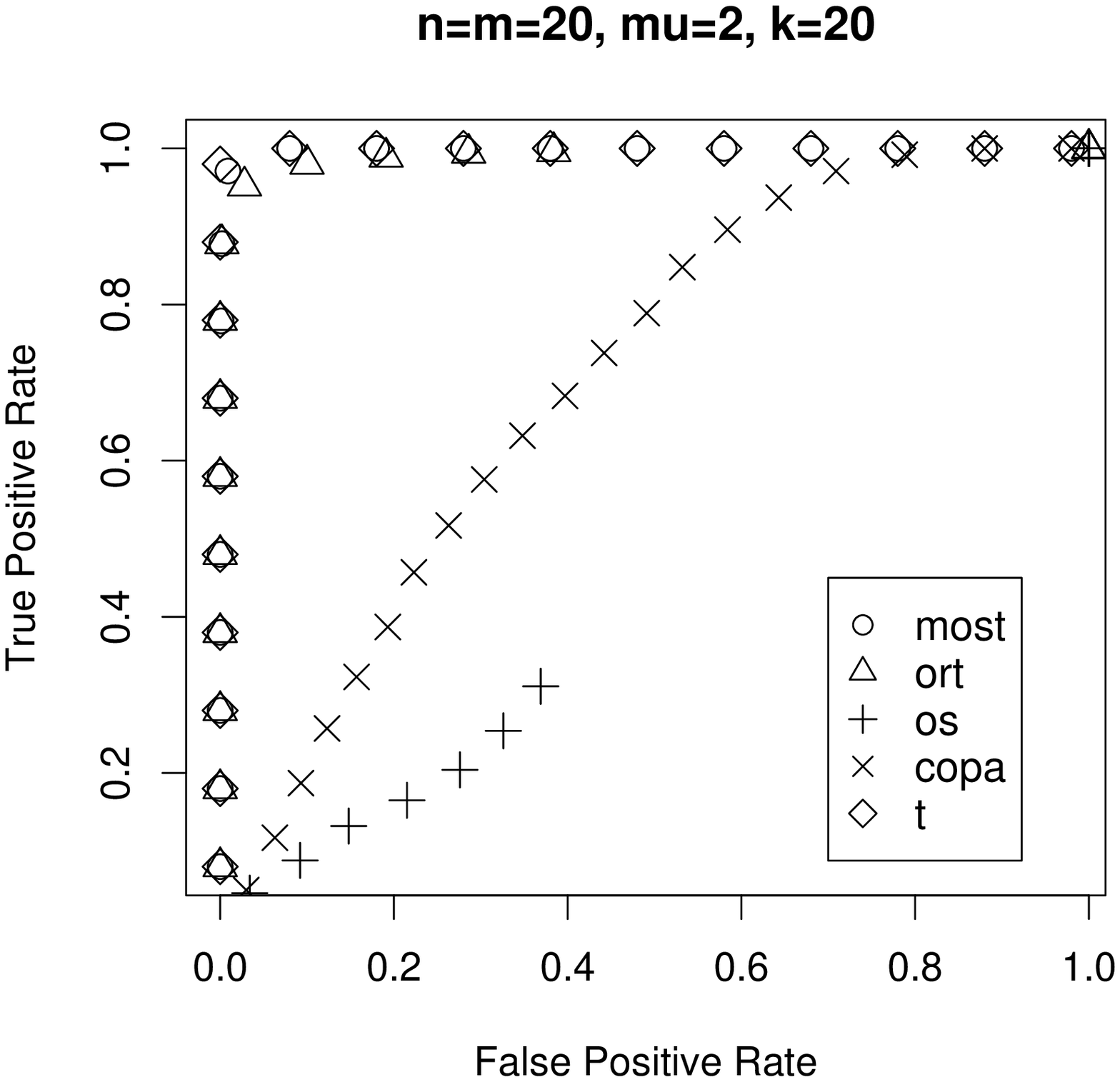} &
\includegraphics[width=6.5cm]{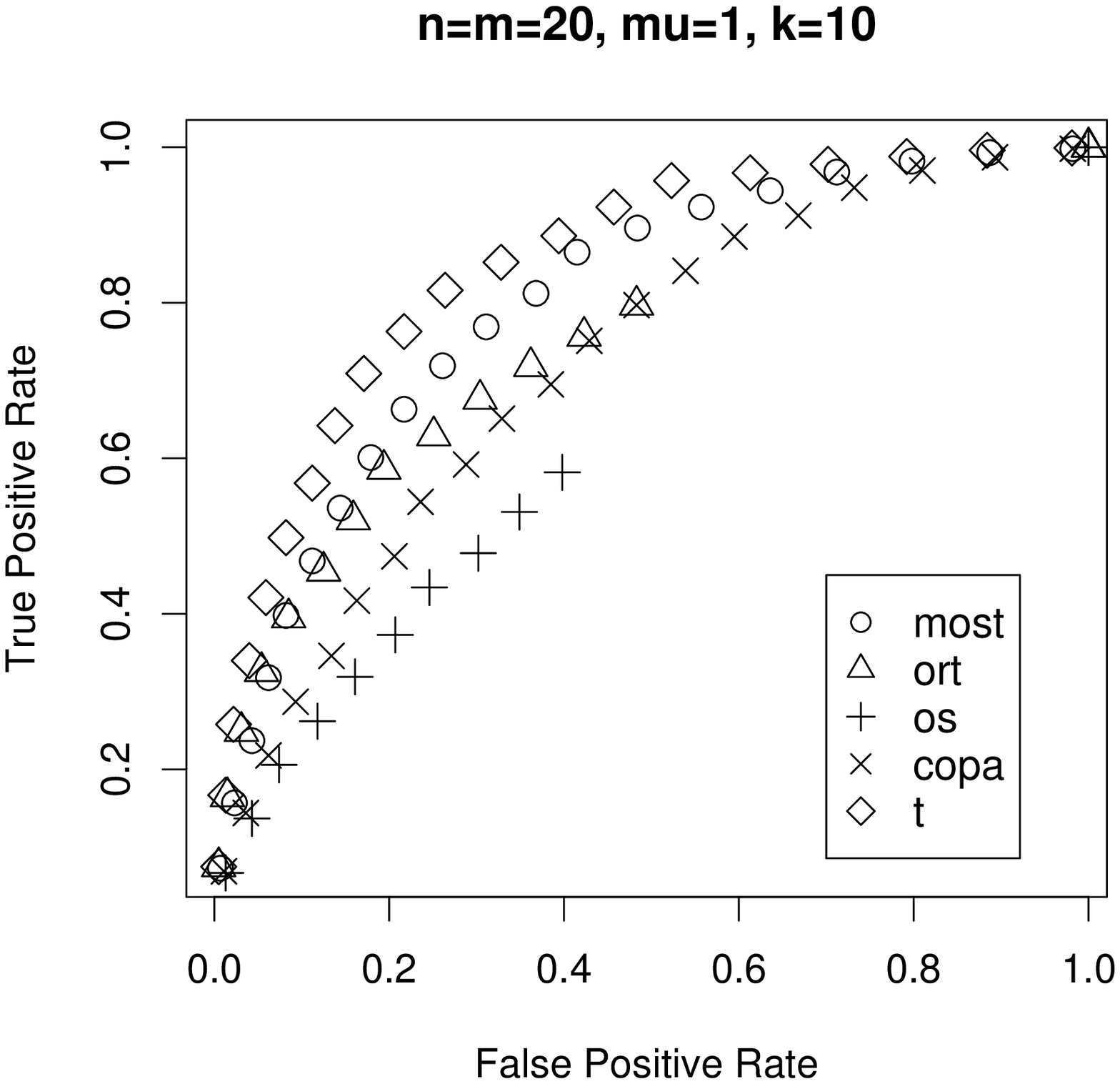} \\
\end{tabular}
\caption{\label{roc1}ROC curves estimated based on simulation. The number of normal/cancer sample is $n=m=20$. Various combinations of $\mu$ and $k$'s are chosen. Other uninteresting results where all statistics have close to perfectly good or bad performances are excluded as explained in the main text.}
\end{figure}

\begin{figure}
\centering
\begin{tabular}{cc}
\includegraphics[width=6.5cm]{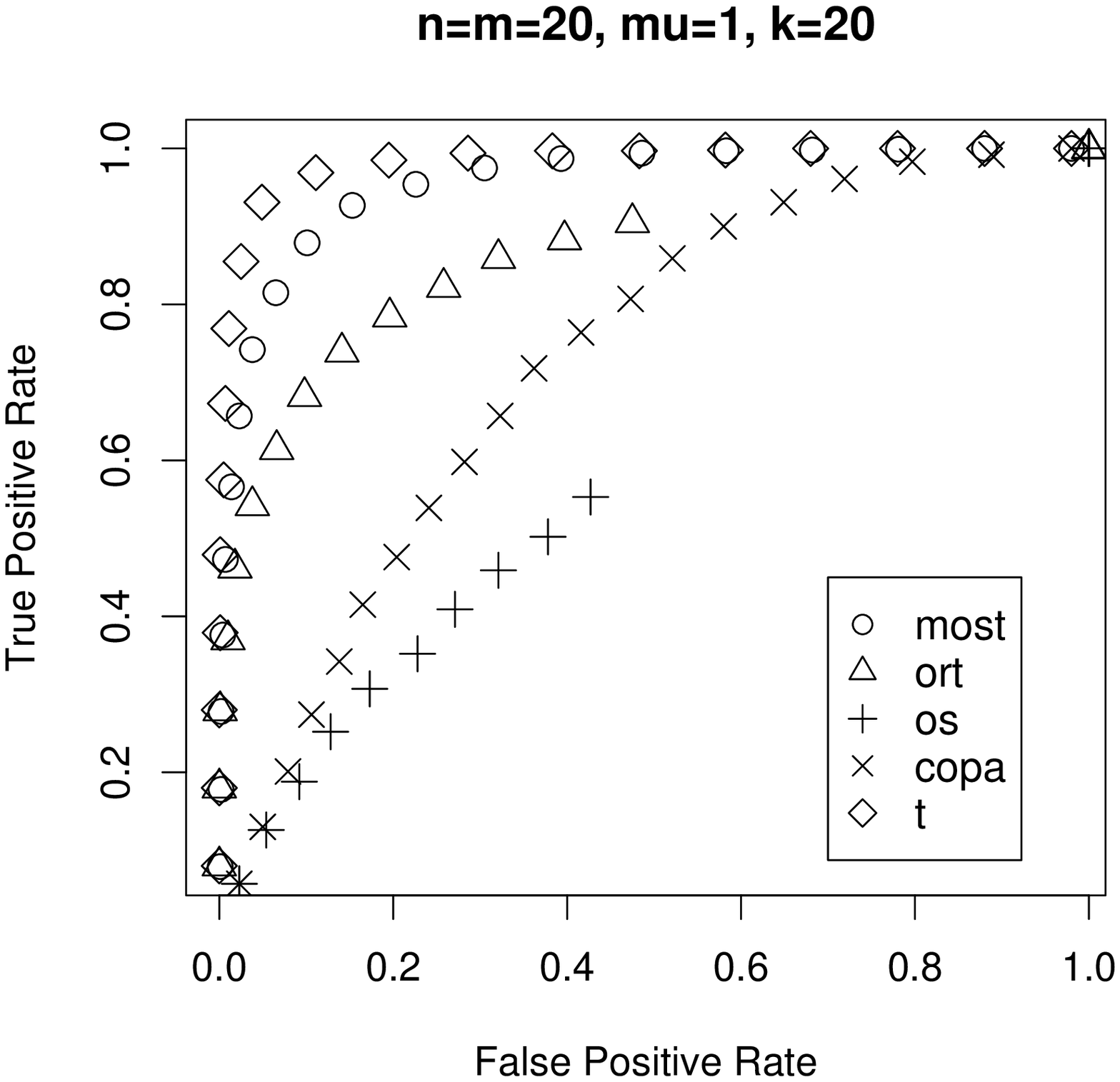} &
\includegraphics[width=6.5cm]{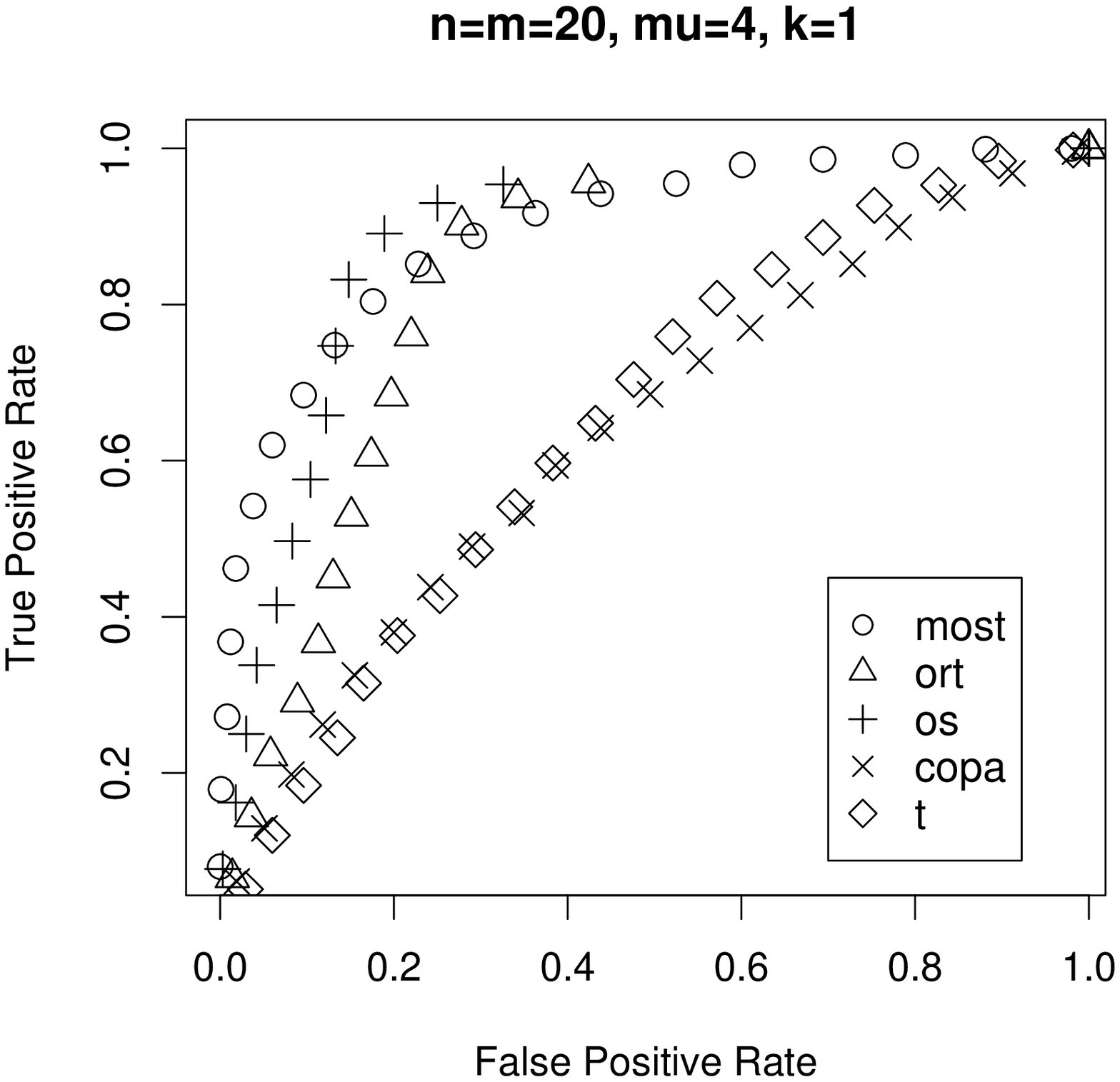} \\
\includegraphics[width=6.5cm]{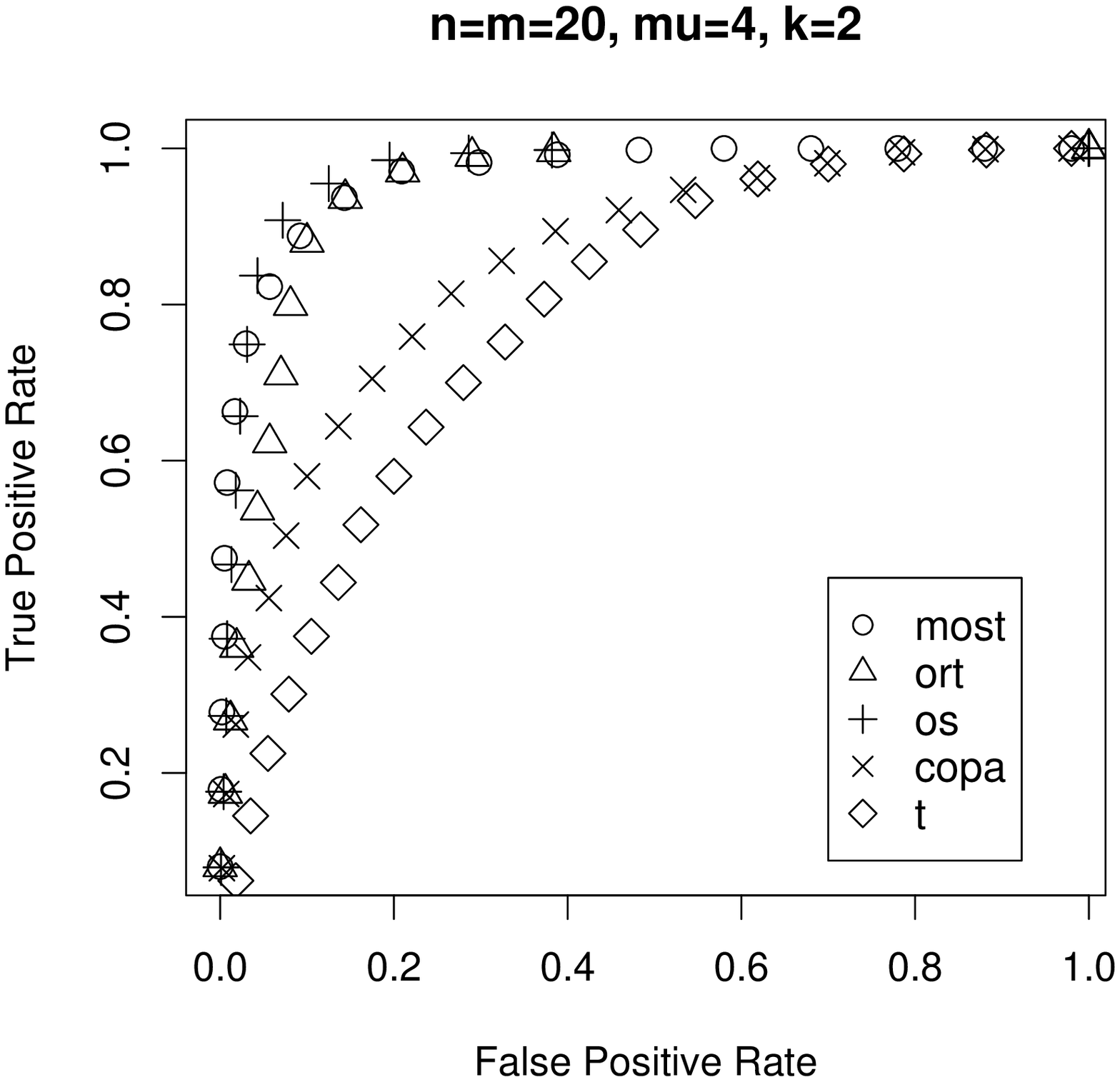} &
\includegraphics[width=6.5cm]{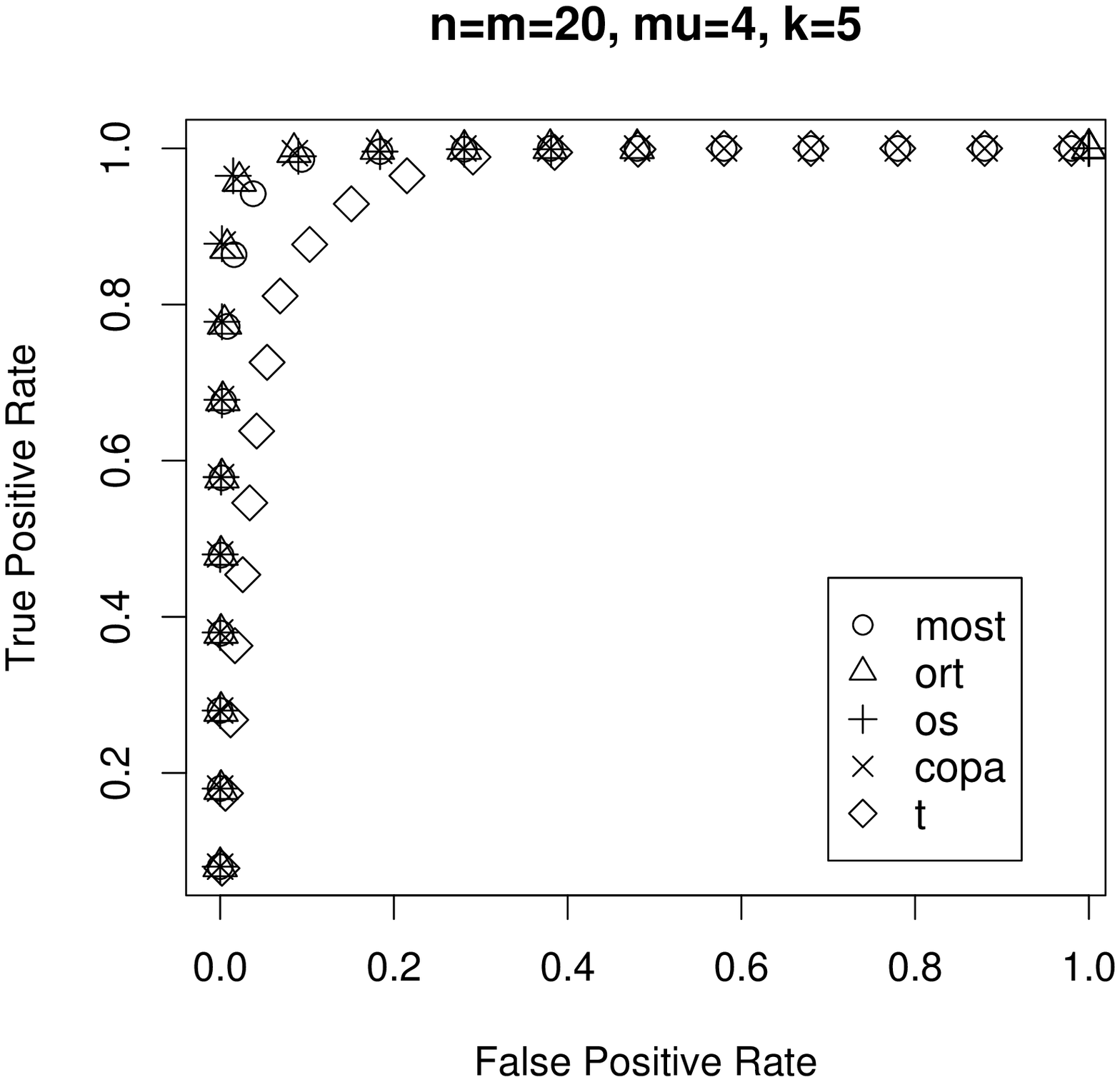} \\

\end{tabular}
\caption{\label{roc2}More ROC curves.}
\end{figure}

\section{Simulation studies and application}
Some simulations are carried out to study MOST, and compare its performance to OS, ORT, COPA, and t-statistics. For COPA, we choose to use the $90$th percentile in its definition as in \cite{tib2006}. We generate the expression data from standard normal with $n=m=20$. For various values $k, 1\le k\le m$, which is the number of differentially expressed cancer samples, a constant $\mu$ is added for differentially expressed genes. We simulated $1000$ differentially and non-differentially expressed genes, and calculated the ROC curves from them by choosing different thresholds for gene calls.

Figure \ref{roc1} and \ref{roc2} plots the ROC curves for some combinations of $k$ and $\mu$. For $\mu=2$ and $k$ small, all five statistics behave similarly with t-statistics performing the worst. As $k$ increases, t becomes better and OS and COPA begin to lose power. For $\mu=1$ and medium to large $k$, the performance of MOST is only worse than t and better than other statistics. Smaller $k$ in this case basically leads to ROC curve that is close to a $45^o$ line for all statistics since the signal $\mu=1$ is too weak in this case, so we do not show these results. For $\mu=4$ and small $k$, MOST is better than ORT, COPA and t, and in this situation only OS is competitive with MOST. Larger $k$ in this case will produce nearly perfect ROC curves  for all statistics, and thus those results are also omitted. Besides ROC curves, we have also tried examining the possibility of using (\ref{k}) for estimating the number of differentially expressed samples $k$, but so far have been unable to get a reasonable estimate out of it. 

From the above simulations, we judge that our new estimate MOST is at least as good as other previously proposed statistics, sometimes much better. Thus it is a valuable tool for detecting activated genes in many situations.

\begin{figure}
\centering
\includegraphics[width=8.5cm]{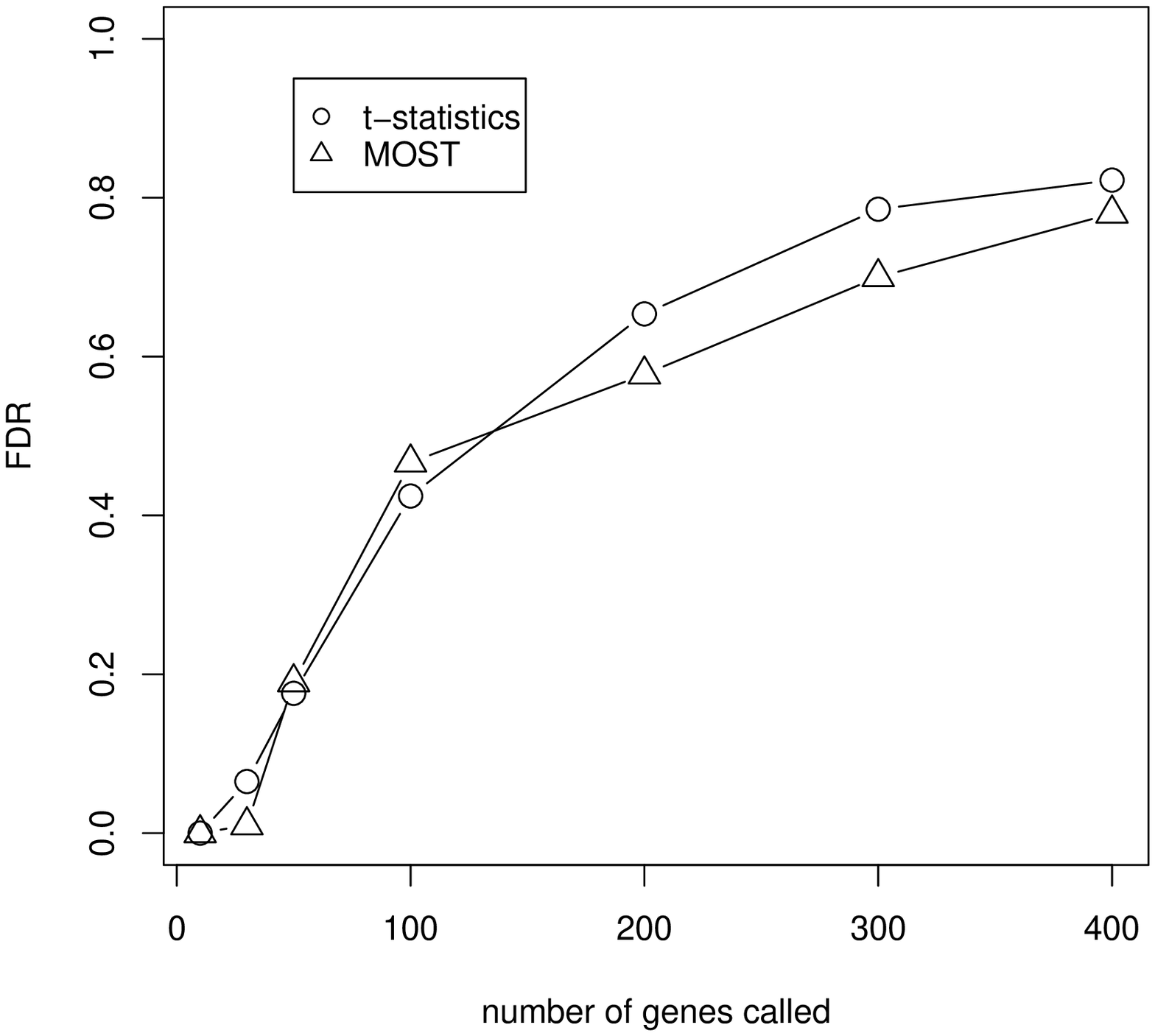} 
\caption{\label{fdr}FDR versus the number of genes called significant.}
\end{figure}

As an example of real data application, the data from \cite{west2001} is publicly available from http://data.cgt.duke.edu/west.php. The microarray used in the breast cancer study contains 7129 genes and 49 tumor samples, 25 of which with no positive lymph nodes identified and the other 24 with positive nodes. Similar to \cite{wu2007}, we take the log transformation of the expressions after normalizing the data. We apply MOST to the data and compare it to the t-statistics by computing the FDR using the SAM approach (\cite{TTC01}). Figure \ref{fdr} plots the FDR versus the number of genes called significant. For this example, MOST seems to perform a little better than t-statistics, although the difference is too small to be of any significance.

\end{document}